\documentstyle[prl,twocolumn,aps,epsf]{revtex}
\begin{document}

\title {Constraints on the size of the quark gluon plasma }
\author{Q.H. Zhang}
\address{Physics Department, McGill university, Montreal QC H3A 2T8, Canada}
\vfill
\maketitle

\begin{abstract}
We use simple entropy arguments 
%the picture given by  Lee\cite{Lee} ten years ago 
to estimate the possible size of the QGP at the AGS and the SPS.  
We find that the possibility to form a large volume of QGP at the AGS 
or the SPS is very small.  The size of the QGP at RHIC and the LHC is also 
predicted.
\end{abstract}

PACS number(s): 25.70 -q, 25.70 Pq,25.70 Gh

%\section{Introduction}

One of the main goals of high energy nuclear collisions 
is to create a quark-gluon plasma (QGP)\cite{Wong1,Hwa1} of macroscopic
size. 
The hope of discovering the QGP in heavy-ion collisions is thus to some 
extent connected to the possibility of measuring the geometric size 
of the region of secondary particle production. 
An important tool in accomplishing such measurements 
is Hanbury Brown-Twiss (HBT) interferometry\cite{Rev,Thmas}.
On the theoretical front, 
owing to the complexity of high energy heavy-ion collisions, 
several models to simulate the 
heavy-ion collisions process\cite{ARC,RQMD,Werner,Wang,ST,others} have
been devised. In those, the size of the QGP phase is a vital element,
as it may affect significantly the shape of the single parton
distributions. 
%To simulate the initial state of QGP one needs to know 
%the size of QGP which may have great influences on the 
%final particle spectrum distribution.  
In this note, using 
HBT results 
from the AGS and the SPS, we infer the possible sizes of QGP in 
those energy regimes.
The sizes of QGP at RHIC and LHC are also 
predicted.

The idea to predict the size of QGP is simple 
and is based on the picture given by Lee some years ago\cite{Lee}.
From the second law of thermodynamics we have 
equation
\begin{equation}
S_{QGP}\le S_{Had}.
\label{e1}
\end{equation}
Here $S_{QGP}$ and $S_{Had}$ are the total entropy 
of the QGP phase and of the hadron phase, respectively. 
As more than eighty percent of final particles are pions, 
we will calculate the entropy of pions and 
 multiply it by a factor of $\alpha$ 
to represent the total particles entropy. For this discussion we 
take $\alpha \sim 1.1$. Eq.(\ref{e1}) can be 
re-written as 
\begin{equation}
s_{QGP}V_{QGP}\le \alpha s_{\pi}V_{\pi}.
\label{e2}
\end{equation}  
Here $V_{QGP}$ and $V_{\pi}$ are the volume of 
the QGP phase and of the pion phase respectively, 
 $s_{QGP}$ and $s_{\pi}$ represent the entropy density 
in the QGP phase and the pion phase, and reads
\begin{eqnarray}
s_{QGP}&=&\frac{2}{45}\cdot\pi^2T^3[16+\frac{21}{2}n_f],
\nonumber\\
s_{Had}&=&\frac{2}{15}\pi^2T^3.
\label{e3}
\end{eqnarray}
$n_f$ is the number of flavors. In the above 
we assume that the pion, quark and antiquark 
masses are zero.  From Eq. (3), we have 
\begin{equation}
\frac{s_{QGP}}{s_{had}}=\frac{16}{3}+\frac{7}{2}n_f= 12\sim 16.
\end{equation}
Using Eq.(2) and Eq.(4), we get 
\begin{equation}
V_{had}\ge 12 V_{QGP}.
\label{e5}
\end{equation}
Gaussian functions have been used 
to fit the two-pion correlation functions at AGS and SPS energy and 
it has been found that the fitted source radius are $5-7 fm$\cite{Rev}. 
From Eq.(\ref{e5}),  we find 
that the source radius for QGP is 
$2-3fm$. This result is important but depends 
strongly on the source radius as measured from pion interferometry. 
If the QGP size is truly $2-3$ fm then we need to consider 
finite size effects on the spectrum distributions. For example, 
because of the 
Heisenberg uncertainty principle, particles in a small volume will 
have a widespread distribution in momentum space. 
It has been shown in Ref.\cite{Wong2,ZP,AS,India} that 
for quarks and gluons, finite size 
effects will become important when the size of QGP is less than 6 fm.
For pions, finite size effects set in when the 
size of the hadron phase is less than 9 fm.  For such small QGP sizes at 
AGS and SPS energy one needs to consider finite size 
effects on the 
final state observables, and this fact is bound to affect several of
the proposed signatures.  
 Finite size effects on the observables 
 have also been studied by 
Elze, Greiner and Rafelski\cite{Frankfurt,Frankfurt1} long ago.
 If we consider the fact that  
pion interferometry actually measures the 
pion source at freezeout time and the freezeout temperature ($T_f$) is
 less than the 
critical temperature ($T_c$), then the QGP source radius 
should be even smaller than the value given above. Using 
typical freezeout temperature $T_f=m_{\pi}=138$ MeV or 
fitted results $T_f = 120$ MeV \cite{density} and assuming  
 the phase transition temperature $T_{c}=200$ MeV, we have
\begin{equation}
V_{had}\ge 37\sim 56 \, V_{QGP}.
\label{e6}
\end{equation} 
This result depends strongly on the freezeout temperature. 
From Eq. (6), we find that the 
 size of QGP phase is $1-2$ fm. This conclusion 
implies that even if a QGP is formed at the AGS or the SPS,  the size of 
QGP will be 
very small. Here we also need to mention that this 
QGP means the QGP at the critical time, {\it i.e.} when the system
enters the mixed phase.
%One can imagine that the size of QGP before 
%critical time should be much 
%more small.  
Of course at the freezeout time, the assumption 
that the pion mass is zero is inappropriate. We will 
come to this question at the end of the paper.
It has been suggested that there is a 
possibility to form some QGP droplets inside a hadron gas.
If this picture is true, we have 
\begin{equation}
S_{had}\ge N_{droplet}S_{droplet}.
\end{equation}
Here $S_{droplet}$ is the entropy in each droplet and 
$N_{droplet}$ is the total number of droplets. 
Assuming that the size of all droplets are the same, we have 
\begin{equation}
V_{had}\ge (37\sim 56) N_{droplet} V_{droplet}.
\end{equation}
If we assume for example that the droplet number is two, the droplet 
size will be 0.5-1.2 fm. 
%That means we will have two 
% bright points (high energy density) in AGS or SPS collision 
%analyses.
%If the above prediction about QGP size is right then lots of 
% signal of QGP which based on QGP picture will be 
%inappropriate as some of those 
%explanations depend on the assumption that 
%the size of QGP is very large. 

The radius of pion sources at RHIC or the LHC are proportional to the pion 
multiplicity distribution as\cite{Satz}: 
\begin{equation}
R_{had}\propto  (0.6\sim 1.2)(\frac{dN}{dy})^{1/2 \sim 1/3}.
\end{equation}
From Eq.(9), we find that 
the hadron phase radius for Au + Au collisions at RHIC 
is 14-26 fm. First HBT results from 
RHIC indicate that the HBT radius is larger 
than the HBT radius at SPS energy, but still 
less than ten Fermis\cite{RHIC}. This may due to the strong flow 
that has been observed at RHIC which make the HBT 
radius less than the true size of hadron phase, $R_{had}$.  
From our numbers for the size of the hadron phase at RHIC we get an
appreciable QGP size of  
\begin{equation}
R_{QGP}=4-8 fm.
\end{equation}
However 
 this estimation depends strongly 
 on the size of hadron phase. This implies a caveat which we will
discuss at the end of this paragraph.   We have assumed that at SPS 
energy, the size of hadron phase is 
 almost the same scale as that of the HBT radius, and this assumption
is borne out by numerical simulations.  
Pion interferometry results from 
simulator (ARC, RQMD, VENUS) at SPS energy  have  
indicated\cite{CGZ} that pion interferometry 
gives us the 
right geometry size of the pion source. A theoretical 
analysis in Ref.\cite{TWU} has shown
that if we use a box source to fit the pion 
correlation function of Pb + Pb collisions at SPS energy, the box 
length is 12 fm, then QGP size at SPS energy will be 
3-4 fm which is still very small.  So both theory and 
empirical analyses indicate that even if a QGP is formed 
at the SPS, it's size should be very small and it is mandatory to consider 
finite size effects.
But at RHIC energy the correspondence between physical size and HBT
radius 
may not hold anymore, owing to strong flow effects.
Present HBT theory shows that HBT only measures  
 part of the whole pion source  due to flow which generates a strong 
correlation between 
coordinate and momentum. This causes the apparent source 
radius to become smaller. 

At LHC energy, the pion multiplicity becomes 
large\cite{Satz} and we estimate 
the radius of the hadronic  phase as
\begin{equation}
R_{had}=18-36 \ {\rm fm},
\end{equation}
and the corresponding QGP size should be 
\begin{equation}
R_{QGP}=6-11 \ {\rm fm}.
\end{equation}

In the following  we will use the Bjorken picture to perform an
independent estimate of QGP size. 
In the Bjorken model, QGP is assumed to be 
produced at first and at time $\tau_0$ it reaches equilibrium 
then the QGP will evolve 
according to hydrodynamical equation, that is\cite{Wong1,Bjorken}
\begin{equation}
s(\tau)\tau=s(\tau_0)\tau_0~~~~~~           (\tau > \tau_0).
\end{equation} 
When the temperature drops to $T_c$ at proper time $\tau_c$, a 
mixed phase occurs, then following equation exists
\begin{equation}
\frac{s(\tau)}{s(\tau_c)}=(\frac{\tau_c}{\tau})^{4/3}~~~~~~ (\tau>\tau_c).
\end{equation}
According to Ref.\cite{Wong1}, after time
\begin{equation}
\tau_h=6.16\tau_c
\end{equation}
the whole system will be in a hadronic phase. 
This phase will expand and temperature
will decrease from critical temperature ($T_c$) to 
freezeout temperature ($T_f$). 
 During the hadron expansion stage, we have 
\begin{equation}
s(\tau_f)\tau_f=s(\tau_h)\tau_h~~~~~(\tau>\tau_h).
\end{equation}
Here $\tau_f$ is the freezeout time. The freezeout time can be 
determined by the following equation\cite{Wong1,Bjorken}
\begin{equation}
\frac{T(\tau)}{T(\tau_h)}=(\frac{\tau_h}{\tau})^{1/3}.
\end{equation}
For $T_f=120$ MeV and $T(\tau_h)=200$  MeV, we have 
\begin{equation}
s(\tau_f)=s(\tau_h)\tau_h/\tau_f=s(\tau_c)(\frac{\tau_c}{\tau_h})^{4/3}
(\frac{T(\tau_f)}{T(\tau_h)})^3.
\end{equation}
Finally we get
\begin{equation}
s(\tau_f)\sim \frac{s(\tau_c)}{52},
\end{equation}
thus 
\begin{equation}
V(\tau_c)=\frac{V_{freezeout}}{52}.
\end{equation}
From Eq.(20) we get similar 
conclusion as above:  {\it even if there is 
QGP formed at AGS or SPS energy, its size should be around the 
size of nucleon}. 

In the following we will use  energy density of pions calculated 
from present measurement to estimate  
the possible QGP energy density at the critical 
time.  Using the Bjorken model and assuming that pion 
mass is zero,  we have 
\begin{equation}
\frac{\epsilon(\tau_h)}{\epsilon(\tau_f)}=(\frac{\tau_f}{\tau_h})^{4/3}.
\end{equation}
Here $\tau_h$ and $\tau_f$ represent hadron time and freezeout time 
respectively. In the mixed phase, Eq.(21) is still valid, implying 
\begin{equation}
\frac{\epsilon(\tau_c)}{\epsilon(\tau_h)}=(\frac{\tau_h}{\tau_c})^{4/3}~~.
\end{equation}
Using $\tau_h=6.16\tau_c$ and assuming (see Eq. (17))
\begin{equation}
\frac{\tau_h}{\tau_f}=(\frac{T(\tau_f)}{T(\tau_h)})^3,
\end{equation}
we have
\begin{equation}
\epsilon(\tau_c)=11.3\cdot \epsilon(\tau_f)(\frac{T(\tau_h)}{T(\tau_f)})^4
	=87\epsilon(\tau_f).
\end{equation}
When $T_f=120$ MeV, it is calculated that the energy density of 
pions
($\pi^+,\pi^-$ and $\pi^0$) is 
 0.023 GeV/fm$^3$. According to Eq.(24), we find that 
the critical density of QGP is  
\begin{equation}
\epsilon(\tau_c) =2.1 {\rm GeV/fm}^3.
\end{equation}
Using equation(with $n_f =2$)
\begin{equation}
\epsilon(\tau_c)=37\frac{\pi^2}{30}T_c^4,
\end{equation}
we find that the critical temperature $T_c =191$ MeV, which is 
consistent with the input value $~T_c =200$ MeV. As we have said earlier 
the assumption of massless pions  
is inappropriate at the freezeout time, as the freezeout 
temperature has almost the same value as the pion mass. 
It has shown in Ref.\cite{density} that pion data from all 
heavy-ion reaction are consistent with thermal 
Bose-Einstein distribution
\begin{equation}
f=\frac{1}{\exp(E/T)-1}
\label{e27}
\end{equation}
with $T=120 MeV$.  Taking the pion mass as 138 MeV, and assuming the 
freezeout temperature $T_f=120 MeV$, the entropy density for 
pions reads
\begin{equation}
s=3\int\frac{d{\bf p}}{(2\pi)^3}[(1+f)ln(1+f)-f\cdot lnf]=
\frac{0.246}{(fm)^3}.
\label{e28}
\end{equation}
Assuming the pion mass as zero, we have $s=0.297/(fm)^3$. 
 That is the entropy density will decrease if 
the value of the mass 
increases.
 Due to  Eq. (28), Eq. (\ref{e6}) changes to 
\begin{equation}
V_{had}\ge 69 V_{QGP}.
\end{equation}  
Thus the QGP size will be even smaller than in our previous estimates.

Let us end with the following comments:

(1) In the above we have assumed that the chemical potentials  
of quarks and pions are zero which is a good approximation if the number 
of quarks and pions are very large. From Fig. 1 or Eq.(\ref{e28}), we 
find that if the chemical potential decreases the entropy density of 
pions will decrease too.
When chemical potential $\mu=m_{\pi}$, the maximum value of pion 
chemical potential, we need to consider Bose-Einstein condensate,
this is of course beyond the scope of the present work. From Fig.1 we 
find that the corresponding maximum entropy density of pions  
is around $0.6/(fm)^3$. Thus we have $V_{had}>27V_{QGP}$ this is the 
upper limit of our estimation. 
%%%%%%%%%%%%%%%%%%%%%%%%%%%%%%%%%%%%%
\begin{figure}[h]\epsfxsize=6cm
\centerline{\epsfbox{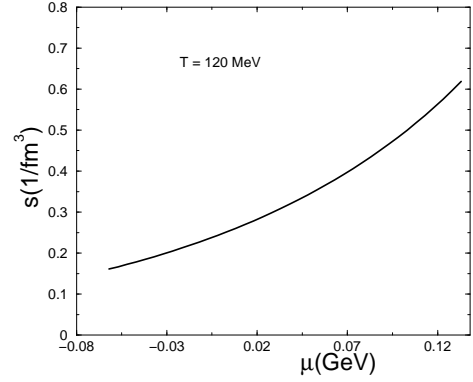}}
\caption{\it Entropy density of pions, $s$, vs. chemical potential, 
$\mu$. In the calculation we fixed the temperature $T=120$ MeV.}
\end{figure} 

(2) In the above calculations, we have used a simple  
Bose-Einstein or Fermi-Dirac distribution for pions or quarks. 
 In principle one needs to use 
 the complete source distribution 
$S(x,K)$ given in Ref.\cite{TWU,Seattle} 
 to calculate the entropy density which will be used to 
estimate the possible QGP 
size at SPS or AGS.  To use the function of $S(x,K)$ or 
$f(x,K)$ to calculate the entropy is very important 
especially when $x$ and $p$ correlation is essential. 
 In Ref\cite{Seattle}, the authors have found that 
\begin{equation}
\frac{S}{N_{\pi}}=3.9\pm 1.8.
\end{equation}
Taking $N_{\pi}=100\sim 150$, we find that 
the QGP size is $2-3fm$, which is consistent with 
the estimations  given above.  We can use 
\begin{equation}
\frac{S}{N_{\pi}}=\frac{\int d{\bf p}[(1+f)ln(1+f)-fln(f)]}
{\int d{\bf p} f}
\end{equation}
to calculate the specific entropy, $S/N$. Using Eq.(\ref{e27}) and 
choosing $T_f =120$ MeV, we 
find that  $\frac{S}{N}=4.32$. This result is consistent 
with Eq. (30).

(3) In this paper,  Eq.(\ref{e3}) has been used to 
calculate the entropy density of QGP phase in finite volume. This formula 
certainly should be corrected when QGP are confined in a finite 
volume. But this is precisely our conclusion: one 
will not observe a large extent QGP at AGS and SPS energies and one 
needs to take into account the finite size.  
 In the following, we
will calculate de Broglie wavelength of quarks, gluons and 
pions to show why we need to take into account 
finite size effects. Using Eq. (\ref{e27}) and taking $T_f=120$ MeV, we 
find that the average pion momentum is 372 MeV. 
%$\langle p\rangle_{\pi} \sim 372 MeV$. 
When $T_c=200$ MeV, the average momentum 
of gluons is 540 MeV 
and the average momentum of quarks is 472 MeV. 
%$\langle p\rangle_q$ is $472 MeV$. 
The de Broglie wavelength for 
pions, quarks and gluons are 
$\lambda_{\pi}=3.3 fm$, $\lambda_g=2.3 fm$ and $\lambda_q=2.6 fm$ respectively. 
It is interesting to notice that the 
de Broglie wavelength for pion at AGS or SPS energy 
is still smaller than the size of pion source. 
But where the de Broglie wavelength for quarks and gluons is almost 
the same size as the size of QGP source, thus finite size effects 
for those particles should be important.  
 
(4) In the paper,  the total entropy 
of final state particles is determined by 
the volume and entropy density of pions.  
But there are also other ways to calculate the final state entropy. 
For example, Bialas and Czyz\cite{BC} have suggested to 
use Renyi's entropy to derive the standard entropy 
$S=-\sum_{n} p_n \log p_n$. Here $p_n$ is the probability of occurrence of 
the state labeled by $n$ and the sum runs over all states 
of the system.  Different from the method given in the paper, 
this method does not explicitly depend on the final state volume. 
One can also calculate the final state entropy by the 
Landau assumption $S=4N$ (consistent with Eq.(30)). 
Here $N$ is the total final state 
particles. This calculation also has nothing to do with 
the final state volume.  Thus, those 
methods are recommended to 
calculate the final state entropy and to infer the possible 
size of the initial state.

(5) All the estimations in this work are based on the assumption 
that the initial state is QGP. If the initial state is a hot pion gas,
a quark-dominated gas, or a gluon-dominated gas, the 
degrees of freedom change and the 
volume of the initial state will change too, for a given final state. 
For example for a hot pion gas, $V_{final}> 5 V_{initial}$.
For a quark gas, $V_{final}>35 V_{inital}$.
For a gluon gas, $V_{final}>26 V_{inital}$. So the initial volume 
is large in the case of a pion gas, as it should be.  

To conclude: it is a key question to know the 
size of QGP which may be produced at the early time of heavy-ion 
collisions.  We show that if there is a QGP formed at the AGS 
and/or at the SPS, its radius should only be around $1-2$ fm. 
Thus the finite size effects on the QGP distribution functions 
will be important. 
%That means previous physical explanation 
%which based on the assumption that infinite or very 
%large QGP will be inappropriate. 
We also predict that the size of QGP at RHIC and LHC 
will be around $4-8$ fm and $6-11 fm$ respectively.  For this size of QGP, 
the finite volume effects on QGP distribution can be neglected.

%%%%%%%%%%%%%%%%%%%%%%%%%%%%%%%%%%%%%
%\begin{figure}[h]\epsfxsize=8cm
%\centerline{\epsfbox{x2.eps}}
%\vskip -2.5cm
%\caption{\it Coherent parameter vs.$ \sigma, R, \Delta $ and 
%$R\Delta$.  Here the pion 
%multiplicity is $100$ and the averaged transverse momentum 
%is 100 MeV. The solid and dashed lines
%correspond to $\lambda $ vs $\sigma$ and $R\Delta$ respectively.}
%\end{figure} 
%%%%%%%%%%%%%%%%%%%%%%%%%%%%%%%%%%%%%%%%%%%%%%%%
\begin{center}
{\bf Acknowledgments}
\end{center}
The author thanks C. Gale for discussions and suggestions. 
This work was partly supported by the 
Natural Sciences and Engineering Research Council of Canada and 
the Fonds FCAR of the Quebec Government.


\begin{thebibliography}{29}
\bibitem{Wong1}
C.Y. Wong, Introduction to high energy heavy-ion collisions,
(World Scientific, Singapore, 1994).
\bibitem{Hwa1}
{\it Quark-Gluon-Plasma 2}, edited by R.C. Hwa ( World Scientific,
Singapore, 1995); U. Heinz and M. Jacob, nucl-th/0002042. 
%\bibitem{Zajc1}
%W. A. Zajc, in Hadronic Multiparticle Production, ed. P. Carruthers
\bibitem{Rev}
Urs. Wiedemann and U. Heinz, Phys. Rept {\bf 319}, 145 (1999).
\bibitem{Thmas}
T. Csorgo, hep-ph/0001233.
\bibitem{ARC}
Y. Pang, T.J. Schlagel, and S.H. Kahana, Phys. Rev. Lett.{\bf 68},2743 (1992).
\bibitem{RQMD}
H. Sorge, H. Stocker, and W. Greiner, Annals. Phys. {\bf 192}, 266 (1989).
\bibitem{Werner}
K. Werner, Phys. Rept. {\bf 232}, 87 (1993).
\bibitem{Wang}
X. N. Wang, Phys. Rept. {\bf 280}, 287 (1997);
M. Gyulassy and X. N. Wang, Comput. Phys. Commun. {\bf 83} 307 (1994).
\bibitem{ST}
B.H. Sa and A. Tai, Comput. Phys. Commun. {\bf 90}, 121 (1995).
\bibitem{others}
B. Zhang, Comput. Phys. Communication {\bf 109}, 193 (1998);
S. Jeon, J. Kapusta, Phys. Rev. C{\bf 56}, 468 (1997).
\bibitem{Lee}
T. D. Lee, Nucl. Phys. A{\bf 538}, 3c (1992).
\bibitem{Wong2}
C. Y. Wong, Phys. Rev. C{\bf 43}, 902 (1993);
M. Mostafa and C. Y. Wong, Phys. Rev. C{\bf 51}, 2135 (1995).
\bibitem{ZP}
Q. H. Zhang and S. Padula, Phys. Rev. C {\bf 62} 024902 (2000).
\bibitem{AS}
A. Ayala and A. Smerzi, Phys. Lett. {\bf B405}, 20 (1997);
A. Ayala, J. Barreiro and L. M. Monta\~no, Phys. Rev. C{\bf 60}, 014904 (1999);
A. Ayala and A. Sanchez, Phys. Rev. C{\bf 63} 064901 (2001).
\bibitem{India}
S. Sarkar, P. K. Roy, D. K. Srivastava and B. Sinha, J. Phys. G{\bf 22}, 951 
(1996).
\bibitem{Frankfurt}
H.-Th. Elze, W. Greiner, J. Rafelski, 
Phys. Lett. {\bf B124}, 515 (1983);
Z. Phys. {\bf C24}, 361 (1984);
Ch. Derreth, W. Greiner, H.-Th. Elze, J. Rafelski, 
Phys. Rev. {\bf C31}, 1360  (1985).
\bibitem{Frankfurt1}
H.-Th. Elze, W. Greiner, 
Phys. Rev. {\bf A33}, 1879 (1986);
Phys. Lett. {\bf B179}, 385 (1986).
\bibitem{density}
G. Bertsch, Phys. Rev. Lett. {\bf 72}, 2349 (1994); Errata {\bf 77}, 789 (1996);
J. Barrette {\em et.al}, Phys. Rev. Lett. {\bf 78}, 2916 (1997);
D. Ference, U. Heinz, B. Tomasik, U. A. Wiedemann, J. G. Cramer,
Phys. Lett. B{\bf 457}, 347 (1999).
\bibitem{Satz}
H. Satz, Nucl. Phys. A{\bf 544}, 371(1992).
\bibitem{RHIC}
S. Y. Panitkin, nucl-ex/0106018.
\bibitem{CGZ}
W.Q. Chao, C.S. Gao and Q.H. Zhang, Nucl. Phys. A{\bf 573}, 641 (1994);
 S. Pratt {\em et al.}, Nucl. Phys. A{\bf 566}, 103c (1994);
 T. Csorgo, J. Zimanyi, J. Bondorf,H. Heiselberg and S. Pratt,
Phys. Lett. B{\bf 241}, 301 (1990); Z.H. Feng, B. H. Sa, R.H. Wang,
Z. Phys. C{\bf 72}, 133 (1996).
\bibitem{TWU}
B. Tomasik, U. A. Wiedemann and U. Heinz, nucl-th/9907096;
A. Ster, T. Csorgo and B. Lorstad, Nucl. Phys. A{\bf 661}, 419 (1999).
\bibitem{Bjorken}
J. D. Bjorken, Phys. Rev D{\bf 27}, 140 (1983).
\bibitem{Seattle}
D. A. Brown, S. Y. Panitkin and G. Bertsch, nucl-th/0002039.
\bibitem{BC}
A. Bialas and W. Czyz; Phys. Rev. D{\bf 61} 074021 (2000);
%Acta Phys. Polon. B{\bf 31} 687 (2000);
K. Fialkowski and R. Wit, Phys. Rev. D{\bf 62} 114016 (2000).
\end{thebibliography}
\end {document}